\documentclass[prb,aps,showpacs,superscriptaddress,twocolumn]{revtex4}%
\usepackage{epsfig}
\usepackage[10pt]{moresize}
\usepackage{amssymb}
\usepackage{comment}
\usepackage{amsmath}
\usepackage{amscd,epsfig,graphicx,rotate}
\usepackage{color}
\usepackage{epsf}
\usepackage[english]{babel}
\usepackage{mathrsfs}
\usepackage[hypertex]{hyperref}
\usepackage{pifont,bm}
\usepackage{amsfonts}

\begin{document}

\title{Detector-induced backaction on the counting statistics of a double
quantum dot}
\author{Zeng-Zhao Li}
\affiliation{Beijing Computational Science Research Center, Beijing 100084, China}
\author{Chi-Hang Lam}
\affiliation{Department of Applied Physics, Hong Kong Polytechnic University, Hung
Hom, Hong Kong, China}
\author{Ting Yu}
\affiliation{Department of Applied Physics, Hong Kong Polytechnic University, Hung
Hom, Hong Kong, China}
\author{J. Q. You}
\affiliation{Beijing Computational Science Research Center, Beijing 100084, China}

\begin{abstract}
Full counting statistics of electron transport is of fundamental importance
for a deeper understanding of the underlying physical processes in quantum
transport in nanoscale devices. The backaction effect from a detector on the
nanoscale devices is also essential due to its inevitable presence in
experiments. Here we investigate the backaction of a charge detector in the
form of a quantum point contact (QPC) on the counting statistics of a biased
double quantum dot (DQD). We show that this inevitable QPC-induced
backaction can have profound effects on the counting statistics under
certain conditions, e.g., changing the shot noise from being sub-Poissonian
to super-Poissonian, and changing the skewness from being
positive to negative. Also, we show that both Fano
factor and skewness can be either enhanced or
suppressed by increasing the energy difference between two single-dot levels
of the DQD under the detector-induced backaction.
\end{abstract}

\maketitle

Current fluctuations in nanoscale systems
provide key insights into the nature of charge transfer beyond what is
obtainable from a conductance measurement alone (see, e.g., Refs.~1 and~2
for recent reviews). An in-depth understanding, however, may require us to
go beyond the first-order 
and even the second-order current correlation functions
(corresponding to the average current and the shot noise respectively) to
study the full counting statistics\cite{Levitov1993JETP,Levitov1996JMP}
which yields all zero-frequency correlation functions at once. Real-time
detection of the tunneling of individual electrons, an important step
towards experimental measurement of the full counting statistics, has
recently been achieved in various QD systems\cite%
{Rimberg2003Nature,Fujisawa2004APL,BylanderDelsing2005Nature}. In
particular, since its measurement in a single QD for the first time\cite%
{Gustavsson2006PRL}, counting statistics has become an important
experimental tool to examine interaction and coherence effects in nanoscale
systems under out-of-equilibrium conditions\cite%
{Flindt2009PNAS,Gabelli2009PRB,Fricke2010APL,Choi2012APL,Ubbelohde2012}.
More recently, counting statistics was applied 
to characterize correlations in both classical and quantum systems\cite%
{Ivanov2012arXiv}.

However, a pronounced effect known as backaction on the counting statistics
of electron transport\cite{Nazarov2003,Nazarov2003EPJB} is inevitably
introduced during measurements made by even most noninvasive detectors such
as a quantum point contact (QPC)\cite{Sukhorukov2007Nphys}. Very recently,
such backaction has been investigated experimentally in a single QD\cite%
{Sukhorukov2007Nphys,Li-GuopingGuo2012APL}. In contrast to a single QD, a
double quantum dot (DQD)\cite{Wiel2002rmp} involves coherent coupling
between two different dots, and therefore can be used to demonstrate
prominent coherent effects\cite{LambertEmaryNori2010prl}. The counting
statistics for DQDs 
has been studied theoretically\cite{Kiessich2006PRB} and experimentally only
for noise properties\cite{Bartholdh2006PRL,Kiessich2007PRL}. In a DQD measured by a QPC, both the current and the shot noise of the QPC have
been previously investigated\cite%
{Gurvitz1997PRB,Korotkov2001PRB,RuskovKorotkov2003PRB,YoungClerk2010PRL}.
In addition, for a zero-bias DQD, the effect of charge-detector-induced
backaction was studied theoretically\cite{You2010PRB} to explain
experimental observations of inelastic electron tunneling\cite%
{Gustavsson2007PRL}. However, to the best of our knowledge, the impacts of
charge-detector-induced backaction on the full counting statistics in these
QD systems have not yet been studied.

Here we investigate the counting statistics of electron transport through a
biased DQD under measurement by a charge detector. 
We demonstrate that this inevitable backaction can indeed have profound
effects on the counting statistics under certain conditions for the DQD. In
particular, it can change the nature of the shot noise from being
sub-Poissonian to super-Poissonian and also change the
skewness from being positive to negative. Moreover, we show that when the
energy difference between two single-dot levels of the DQD increases, %
both Fano factor and skewness can be
either enhanced or suppressed under the detector-induced backaction. These
QPC-backaction-induced effects are expected to be experimentally observable
with currently existing technologies. Apart from 
a deeper understanding of experimental observations, this study may also
shed light on how to control these QD systems using the backaction of a
charge detector.

\section*{Results}

We focus on a setup consisting of a lateral DQD, which is coupled to the
source and the drain electrodes, and measured by a nearby QPC [see
Figure~1(a)]. The lateral DQD is formed by properly tuning the voltages
applied to the corresponding gates. Here we consider a Coulomb-blockade
regime with strong intradot and interdot Coulomb interactions, so that only
one electron is allowed in the DQD system. The states of the DQD are
represented by the occupation 
states $|1\rangle $ and $|2\rangle $, denoting one electron in the left and
the right dots, respectively [see Figure~1(b)].


The total Hamiltonian of the whole system can be written as
\begin{equation}
H=H_{\mathrm{DQD}}+H_{\mathrm{leads}}+H_{\mathrm{QPC}}+H_{\mathrm{T}%
}+H_{\det },
\end{equation}
where (we set $\hbar =1$)%
\begin{equation}
H_{\mathrm{DQD}}=\frac{\varepsilon }{2}\sigma _{z}+\Omega \sigma _{x},
\end{equation}%
\begin{equation}
H_{\mathrm{leads}}=\sum_{s}(\omega _{ls}c_{ls}^{\dag }c_{ls}+\omega
_{rs}c_{rs}^{\dag }c_{rs}),
\end{equation}%
\begin{equation}
H_{\mathrm{QPC}}=\sum_{kq}(\omega _{Sk}c_{Sk}^{\dag }c_{Sk}+\omega
_{Dq}c_{Dq}^{\dag }c_{Dq}),
\end{equation}%
\begin{equation}
H_{\mathrm{T}}=\sum_{s}[(\Omega _{ls}c_{ls}^{\dag }a_{1}+\Omega
_{rs}\Upsilon _{r}^{\dag }c_{rs}^{\dag }a_{2})+\mathrm{H.c.}],
\end{equation}
\begin{equation}
H_{\mathrm{\det }}=\sum_{kq}(T-\zeta \sigma _{z})(c_{Sk}^{\dag
}c_{Dq}+c_{Dq}^{\dag }c_{Sk}).
\end{equation}
Here, $H_{\mathrm{DQD}}$, $H_{\mathrm{leads}}$, and $H_{\mathrm{QPC}}$ are,
respectively, the free Hamiltonians of the DQD, the electrodes coupled to
the DQD, and the QPC without the tunneling term. In the DQD Hamiltonian, $%
\varepsilon$ is the energy difference between the two single-dot levels and $%
\Omega$ the interdot tunneling-coupling strength. Also, we define pseudospin
operators $\sigma _{z}\equiv a_{2}^{\dag }a_{2}-a_{1}^{\dag }a_{1} $ and $%
\sigma _{x}\equiv a_{2}^{\dag }a_{1}+a_{1}^{\dag }a_{2}$, with $a_{1}$ ($%
a_{2}$) being the annihilation operator for an electron staying at the left
(right) dot. $c_{ls}$ ($c_{rs} $) is the annihilation operator for electrons
in the source (drain) reservoir, i.e., the left (right) electrode of the
DQD, while $c_{Sk}$ ($c_{Dq}$) is the annihilation operator for electrons in
the source (drain) reservoir of the QPC with momentum $k$ ($q$). $H_{\mathrm{%
T}}$ gives the tunneling-coupling Hamiltonian between the DQD and the two
electrodes where the counting operator $\Upsilon _{r}$ ($\Upsilon
_{r}^{\dagger }$) decreases (increases) the number of electrons that have
tunneled into the right electrode (via the barrier between the DQD and the
right electrode)\cite{Doiron07}. These counting operators are introduced to
keep track of the progress of the tunneling processes by successive
electrons. Finally, $H_{\mathrm{\det }}$ describes tunnelings in the QPC
which depends on the electron occupation of the DQD, owing to electrostatic
couplings between the DQD and the QPC. We define $T\equiv T_{0}-(\zeta
_{2}+\zeta _{1})/2$ and $\zeta \equiv (\zeta _{2}-\zeta _{1})/2$, so that
the transition amplitudes of the QPC, when an extra electron staying at the
left and the right dots, equal $T+\zeta $ and $T-\zeta $, respectively\cite%
{You-Li2012PRB}.

\subsection{Counting statistics.}

To study the counting statistics of the electron transport through a DQD system, it
is essential to know the probability $P( n,t)$ of $n$ electrons having been
transported from the DQD to the right electrode during a period of time $t$.
It is related to the cumulant generating function $G(\chi,t)$ defined by\cite%
{Nazarov2003}
\begin{equation}
e^{-G(\chi,t)}=\sum_{n}P(n,t) e^{i\chi n}.  \label{eq:CGF}
\end{equation}
We consider the time interval $t$ much longer than the tunneling time of an
electron through the DQD system, so that transient properties %
(i.e., finite-frequency counting
statistics)\cite{Emary2007,Marcos2010,AlbertFlindtButtiker2011prl} are
insignificant. The derivative of the cumulant generating function with
respect to the counting field $\chi $ at $\chi =0$ yields the $j$-th
cumulant, i.e., $C_{j}=-(-i\partial _{\chi }) ^{j}G(\chi,t)|_{\chi
\rightarrow 0}$, where $\chi $ is a field conjugate to $n$ (see, e.g.,
Ref.~2).
These cumulants carry complete information on the counting statistics of the
DQD system. For instance, the average current and the shot noise can be
expressed as $I=eC_{1}/t$ and $S=2e^{2}C_{2}/t$. Thus, the Fano factor $F$,
which is used to characterize the bunching and anti-bunching phenomena in
the transport processes, is given by $F =S/2eI=C_{2}/C_{1}$. %
The skewness is defined by $K=C_{3}/C_1$, which characterizes the asymmetric degree of the distribution of the transported electrons around its mean
value.

On the other hand, the probability-distribution function of the number of
transported electrons 
can also be expressed as
\begin{equation}
P(n,t)=\rho _{00}^{n}(t)+\rho _{gg}^{n}(t)+\rho _{ee}^{n}(t),
\label{eq:probability}
\end{equation}%
where $\rho _{ij}^{n}(t)$ $(i,j\in \{0,g,e\})$ denote the reduced density
matrix elements of the DQD at a given number $n$ of electrons transported
from the DQD to the right electrode in time $t$. Here $0$, $g$, and $e$
denote the eigenstates $|0\rangle $, $|g\rangle $, and $|e\rangle $ of the
DQD, which correspond to no electron staying in the DQD, one electron in the
ground state, and one electron in the excited state, respectively [see
horizontal solid lines in Figure~1(b)]. From equations~(\ref{eq:CGF}) and~(%
\ref{eq:probability}), we have%
\begin{eqnarray}
G(\chi ,t) &=&-\ln \left\{ \sum\limits_{n}\left[ \rho _{00}^{n}(t)+\rho
_{gg}^{n}(t)+\rho _{ee}^{n}(t)\right] e^{i\chi n}\right\}   \notag \\
&=&-\ln \left[ \rho _{00}(\chi ,t)+\rho _{gg}(\chi ,t)+\rho _{ee}(\chi ,t)%
\right]   \notag \\
&=&-\ln \mathrm{Tr}[\rho _{ij}(\chi ,t)], \label{eq:cgf2}
\end{eqnarray}%
with%
\begin{equation}
\rho _{ij}(\chi ,t)=\sum_{n}\rho _{ij}^{n}(t)e^{i\chi n}.  \label{eq:fourier}
\end{equation}
Note that the reduced density matrix elements $\rho _{ij}^{n}(t)$ in
equation~(\ref{eq:fourier}) satisfy a master equation (see Methods) and $\rho _{ij}(\chi ,t)$ are just the Fourier transforms of these matrix elements\cite{Kiessich2006PRB,Kiessich2007PRL}. Below we manage to
obtain the cumulant generating function $G(\chi ,t)$ at a long time $t.$

Based on the master equation of $\rho _{ij}^{n}(t),$ we can derive the following
equation of motion:
\begin{equation}
\frac{\partial \varrho }{\partial t}=-\mathcal{M}(\chi )\varrho ,
\end{equation}%
with \begin{widetext}
\begin{equation}
\mathcal{M}(\chi )\!=\!\left(
\begin{array}{ccccc}
\Gamma _{L} & -\beta ^{2}\Gamma _{R}e^{i\chi } & -\alpha ^{2}\Gamma
_{R}e^{i\chi } & 2\alpha \beta \Gamma _{R}e^{i\chi } & 0 \\
-\alpha ^{2}\Gamma _{L} & \beta ^{2}\Gamma _{R}\!+\!\gamma _{\mathrm{ex}} &
-\gamma _{re} & -\alpha \beta \Gamma _{R}\!-\!2\eta \gamma _{\mathrm{de}} & 0
\\
-\beta ^{2}\Gamma _{L} & -\gamma _{\mathrm{ex}} & \alpha ^{2}\Gamma
_{R}\!+\!\gamma _{re} & -\alpha \beta \Gamma _{R}\!+\!2\eta \gamma _{de} & 0
\\
-\alpha \beta \Gamma _{L} & -\frac{1}{2}\alpha \beta \Gamma _{R}\!-\!\eta
\gamma _{\mathrm{ex}} & -\frac{1}{2}\alpha \beta \Gamma _{R}\!+\!\eta \gamma
_{\mathrm{re}} & \frac{1}{2}\Gamma _{R}\!+\!2\eta ^{2}\gamma _{\mathrm{de}}
& -\Delta  \\
0 & 0 & 0 & \Delta  & \frac{1}{2}\Gamma _{R}\!+\!2\eta ^{2}\gamma _{\mathrm{%
de}}\!-\!\gamma _{\mathrm{ex}}\!-\!\gamma _{\mathrm{re}}%
\end{array}%
\right) ,
\label{matrix}
\end{equation}%
\end{widetext}
where $\varrho \equiv (\rho _{00}(\chi ,t),\ \rho _{gg}(\chi ,t),\ \rho
_{ee}(\chi ,t),$ $\mathrm{Re}[\rho _{eg}(\chi ,t)]$, $\mathrm{Im}[\rho
_{eg}(\chi ,t)])^{T}$. Note that $\rho_{0g}(\chi ,t)$ and
$\rho_{0e}(\chi ,t)$ as well as their complex conjugates are decoupled from
the reduced density matrix elements given above and therefore are not included. In the matrix $\mathcal{M}%
(\chi )$, $\alpha =\cos ({\theta }/{2})$, $\beta =\sin ({\theta }/{2})$, and
$\eta \!=\!\cos \theta $, with $\tan \theta \!=\!2\Omega /\varepsilon $; $%
\Delta \!=\!\sqrt{\varepsilon ^{2}+4\Omega ^{2}}$, and $\Gamma _{L(R)}=2\pi
g_{L(R)}\Omega _{lk(rk)}^{2}$ is the rate of electron tunneling through the
barrier between the DQD and the left (right) electrode. Here, $g_{i}$ ($i=L$%
, or $R$) denotes the density of states at the left or the right electrode
of the DQD, which is assumed to be constant over the relevant energy range.
The QPC-induced excitation rate $\gamma _{\mathrm{ex}}$, relaxation rate $%
\gamma _{\mathrm{re}}$, and dephasing rate $\gamma _{\mathrm{de}}$ are given
by
\begin{eqnarray}
\gamma _{\mathrm{ex}} &=&\lambda \left[ \Theta \left( eV_{\mathrm{QPC}%
}-\Delta \right) +\Theta \left( -eV_{\mathrm{QPC}}-\Delta \right) \right] \,,
\label{backactionRates1} \\
\gamma _{\mathrm{re}} &=&\lambda \left[ \Theta \left( eV_{\mathrm{QPC}%
}+\Delta \right) +\Theta \left( -eV_{\mathrm{QPC}}+\Delta \right) \right] \,,
\label{backactionRates2} \\
\gamma _{\mathrm{de}} &=&\lambda \left[ \Theta \left( eV_{\mathrm{QPC}%
}\right) +\Theta \left( -eV_{\mathrm{QPC}}\right) \right] \,,
\label{backactionRates3}
\end{eqnarray}%
where $\Theta (x)=(x+|x|)/2$, and $\lambda =2\pi g_{S}g_{D}\zeta ^{2}$, with $g_{S}$ $(g_{D})$ being the
density of states of the source (drain) electrode in the QPC. We also assume
$g_{S}$ and $g_{D}$ to be constant over the relevant energies\cite%
{Strace-Barrett2004PRL}. Let $\Lambda_{\mathrm{min}}$ be the minimal eigenvalues of $\mathcal{M}(\chi )$ in equation (\ref{matrix}). At a long time $t$, the behavior of $\varrho (\chi ,t)$ is dominantly governed by\cite{Bagrets-Nazarov2003PRB,Flindt2005EPL,Flindt2008PRL,Braggio2009JStatMech,Luo2011JPCM}
\begin{equation}
\varrho (\chi ,t)=e^{-\Lambda _{\mathrm{min}}t}\varrho (\chi ,0).
\end{equation}%
Therefore, we have $\rho _{ii}(\chi ,t)=e^{-\Lambda _{\mathrm{min}}t}\rho
_{ii}(\chi ,0)$ at a long time $t$. Then, it follows from equation (\ref{eq:cgf2}) that the cumulant generating function at both a small $\chi $ and a long time $t$ is given by
\begin{equation}
G(\chi ,t)=\Lambda _{\mathrm{min}}t,
\end{equation}
because $\left[ \rho _{00}(\chi ,0)+\rho _{gg}(\chi ,0)+\rho _{ee}(\chi ,0)%
\right] |_{\chi \rightarrow 0}=\sum_{i}\sum_{n}\rho
_{ii}^{n}(0)=\sum_{i}\rho _{ii}(0)=1.$

Note that the dephasing rate given in equation (\ref{backactionRates3}) is proportional to the bias voltage
of the QPC, which is consistent with a previous study\cite{Gurvitz1997PRB}.
In addition, our approach is based on the Born-Markov approximation (see
Methods), which applies when the rates induced by the backaction from the
QPC are weak. With the transition rate of a single electron hopping from one
reservoir of the QPC to the other, we can use the Landauer formula to obtain
the transition probability\cite{Gurvitz1997PRB} and then have $g_{S}g_{D}=%
I_{\mathrm{QPC}}\hbar /(2\pi T^{2}e^{2}V_{\mathrm{QPC%
}})$, where $I_{\mathrm{QPC}}$ and $V_{\mathrm{QPC}}$
are the current and the bias voltage of the QPC with the densities of states
$g_{S}$ and $g_{D}$ at the source and the drain reservoirs. Following a
recent experiment reported in Ref.~35, we take $I_{\mathrm{QPC}%
}=500$~nA and $V_{\mathrm{QPC}}=0.5$~meV to determine $g_{S}g_{D}$. In addition, we choose
$\zeta/T=0.044$ as in Ref.~30, so that the QPC conductance
changes by $\sim 1\%$ if the number of electrons in the DQD changes by one%
\cite{Vandersypen2004APL}.%



\subsection{Detector-induced backaction under resonance condition.}

In order to obtain some compact analytical results for the counting statistics,
we first consider, for simplicity, the resonance case where energy
difference between two single-dot levels is zero (i.e., $\varepsilon =0$).
For instance, the charge current through the DQD is obtained as
\begin{equation}
I=\frac{4e\Gamma _{L}\Gamma _{R}\,\Omega ^{2}}{\Xi },  \label{current}
\end{equation}%
and shot noise is
\begin{equation}
S=\frac{8e^{2}\Gamma _{L}\Gamma _{R}\,\Omega ^{2}f}{\Xi ^{3}},
\label{shot_noise}
\end{equation}%
where%
\begin{equation}
\Xi =\Gamma _{L}\Gamma _{R}^{2}+4\left( 2\Gamma _{L}+\Gamma _{R}\right)
\Omega ^{2}-2\Gamma _{L}\Gamma _{R}\left( \gamma _{\mathrm{ex}}+\gamma _{%
\mathrm{re}}\right) ,  \label{para11}
\end{equation}%
\begin{eqnarray}
f\! &=&\!16\Gamma _{R}^{2}\Omega ^{4}+\Gamma _{L}^{2}\left( \Gamma
_{R}^{4}-8\Gamma _{R}^{2}\Omega ^{2}+64\Omega ^{4}\right) +4\Gamma
_{L}^{2}\Gamma _{R}  \notag \\
&&\times \left( \gamma _{\mathrm{ex}}+\gamma _{\mathrm{re}}\right) \left[
-\Gamma _{R}^{2}-4\Omega ^{2}+\Gamma _{R}\left( \gamma _{\mathrm{ex}}+\gamma
_{\mathrm{re}}\right) \right] .  \label{para12}
\end{eqnarray}%
Then, the Fano factor $F\equiv S/2eI$ also follows straightforwardly.

From equations~(\ref{current})-(\ref{para12}), it is clear that the charge
current $I$, the shot noise $S$, and hence the Fano factor $F$ depend on the
excitation rate $\gamma_{\mathrm{ex}}$ and the relaxation rate $\gamma_{%
\mathrm{re}}$ induced by the QPC. These reveal the impacts of the backaction
from the charge detector. More importantly, because of the nontrivial
dependence of both $\gamma_{\mathrm{ex}}$ and $\gamma_{\mathrm{re}}$ on the
applied voltage across the QPC [see equations~(\ref{backactionRates1}) and~(%
\ref{backactionRates2})], the presence of the charge-detector-induced
backaction can be experimentally checked.
Note that in the case of no backaction, where $\gamma _{\mathrm{ex}}$ and $\gamma
_{\mathrm{re}}$ in equations~(\ref{current}) and (\ref{shot_noise}) are
equal to zero, our results reduce to the previous results obtained by
other approaches\cite{Gurvitz1996PRB,Nazarov1996PRB}. For simplicity, the
temperature is here chosen to be zero because it is extremely low in
quantum-transport experiments. Other parameters like the interdot coupling
strength $\Omega$ and the tunneling rate $\Gamma_{L}$ are taken from the
experimental data \cite{Bartholdh2006PRL,Kiessich2007PRL}.

The charge current obtained from the cumulant $C_{1}$ of the counting
statistics is calculated both {\it with} and {\it without} backaction, and the results
are presented in Figure~2.
When the backaction from the charge detector is taken into account, we
observe that the current through the DQD is significantly enhanced as shown
in Figure~2(a). In particular, when $\left\vert eV_{\mathrm{QPC}}\right\vert
\leq \Delta$, a plateau with a constant current is observed [see, e.g.,
Figure~2(b)]. This plateau corresponds to a regime in which QPC-induced
excitations is suppressed but there is still a constant relaxation rate
contributed by the presence of the QPC, as can be interpreted from
equations~(\ref{backactionRates1}) and~(\ref{backactionRates2}). Physically,
a critical energy $\Delta$ exists for the QPC-induced excitation of an
electron in the DQD and is hence required to change 
the current\cite{You2010PRB,Gustavsson2007PRL}. 
Beyond the regime of constant current, i.e., $\left\vert eV_{\mathrm{QPC}%
}\right\vert > \Delta$, it is clearly shown that in the region shown in Figure~2(b) the current increases with
the magnitude of the 
voltage applied across the QPC in a nearly linear manner.

For the Fano factor $F\equiv S/2eI$,
our results are given in Figure~3. As shown in Figure~3(a), the nature of
the shot noise can be changed from \textit{sub}-Poissonian ($F<1$) to
\textit{super}-Poissonian ($F>1$), and vice versa, under the QPC-induced
backaction. Without backaction, e.g., when the QPC is decoupled to the DQD,
the Fano factor is always smaller than one, i.e., \textit{sub}-Poissonian, implying the anti-bunching of electrons [see the dashed line in
Figure~3(b)]. If QPC-induced backaction is considered but with a condition $%
\left\vert eV_{\mathrm{QPC}}\right\vert \leq \Delta$, we find a 
plateau similar to 
that in the current. 
As mentioned above, the electron transport in this regime does not involve
QPC-induced excitations. Outside this plateau (i.e. $\left\vert eV_{\mathrm{%
QPC}}\right\vert > \Delta$), the Fano factor increases with $\left\vert V_{%
\mathrm{QPC}}\right\vert$. For a sufficiently large bias, we can get $F=1$
[see Figures~3(b) and the solid curve in Figure~3(a)], indicating that the
electron transport is uncorrelated in time and is described by \textit{%
Poissonian} statistics. Beyond this large bias, we have $F >1$.
Thus, bunching of electrons in the transport through the DQD occurs,
resulting in \textit{super}-Poissonian noise. 
%
Physically, the effective tunneling rates for two eigenstate channels are
obtained as $\Gamma_{R}^{\left( g\right) }=\beta ^{2}\Gamma_{R}+\gamma
_{\mathrm{ex}}$ and $\Gamma _{R}^{\left( e\right) }=\alpha^{2}\Gamma _{R}+\gamma
_{\mathrm{re}} $ (see Methods). Without backaction, i.e., $\gamma_{\mathrm{re}}=\gamma_{\mathrm{ex}}=0$%
, it follows that $\Gamma_{R}^{\left( g\right) }=\Gamma _{R}^{\left( e\right) }$ under
the resonance condition (i.e., $\varepsilon=0$) because of $\alpha=\beta$. This corresponds to \textit{sub}%
-Poissonian noise (i.e., $F<1$). For example, $F\approx0.39$ for the dashed line in Figure~3(b). When the detector-induced backaction is included, i.e., $\gamma_{\mathrm{re}},\gamma_{\mathrm{ex}}\neq0$, the effective
tunneling rates become unequal (i.e., $\Gamma _{R}^{\left( e\right)
}\neq\Gamma _{R}^{\left( g\right)}$) even if $\alpha=\beta$ under the resonance condition. This increases $F$ but $F$ is still smaller than $1$ for small values of $\left\vert eV_{\mathrm{QPC}}\right\vert$. However, when $\left\vert eV_{\mathrm{QPC}}\right\vert$ further increases, it enhances $\gamma_{\mathrm{re}}$ and $\gamma_{\mathrm{ex}}$ [see equations~(\ref{backactionRates1}) and~(%
\ref{backactionRates2})], i.e., the relaxation and the excitation. This makes the two effective tunneling rates more asymmetric, yielding $F>1$ (i.e., \textit{super}-Poissonian noise) at large values of $\left\vert eV_{\mathrm{QPC}}\right\vert$. Therefore, the change of shot noise from being \textit{sub}-Poissonian to \textit{super}-Poissonian is due to the effect of dynamical blocked channels\cite{Belzig2004PRL,Belzig2004EPL,Belzig2005PRB} induced by the QPC backaction.

We also numerically calculate the skewness $K=C_3/C_1$ (see Figure~4). As demonstrated in Figure~4(a), the skewness can be changed from being positive ($K>0$) to negative ($K<0$), and vice versa, under the QPC-induced backaction. Here $K=0$ corresponds to a symmetric Gaussian distribution of electron tunneling, where the tunneling of the larger number of electrons in a given time duration occurs with the same probability as the tunneling of the smaller number of electrons, with respect to a mean value. For $K>0$ ($K<0$), the distribution of electron tunneling becomes asymmetric with the tunneling of the larger number of electrons in a given duration occurring with a higher (lower) probability. Without backaction, the skewness is always positive [see the dashed line in Figure~4(b)]. When
QPC-induced backaction is included but with $\left\vert eV_{\mathrm{QPC}}\right\vert \leq \Delta$, a plateau similar to
that in either current or Fano factor appears. In this region, QPC-induced
excitations are not involved in the electron transport. Outside this region
(i.e., $\left\vert eV_{\mathrm{QPC}}\right\vert > \Delta$), the skewness can be either positive or negative, depending on the values of $\Gamma_R$ and $\left\vert eV_{\mathrm{QPC}}\right\vert$ [see Figure~4(b)
and the regions surrounded by solid curves in Figure~4(a)]. This indicates that the distribution of transported electrons can deviate from the Gaussian in an opposite way.

\subsection{Detector-induced backaction under off-resonance condition.}

Note that the QPC-induced dephasing is not discussed above. This is because
such dephasing with a rate $\gamma _{\mathrm{de}}$ ($=\lambda \left\vert eV_{%
\mathrm{QPC}}\right\vert $) does not induce any transitions between
different states of the DQD. However, it is known that dephasing produces
broadening of the energy levels of the DQD\cite{Gurvitz1997PRB}, and then
can affect the current. Indeed, in this case, the resonance condition (i.e.,
$\varepsilon =0$) may be violated. Below we numerically study the backaction
effect under the off-resonance condition (i.e., $\varepsilon \neq 0$)
because analytical results cannot be derived in this more general case.

To investigate the effect of the energy difference $\varepsilon$ on the
detector-induced backaction, we calculate the current, the shot noise, the
Fano factor, and the skewness using the same parameters as
in Figures~2(b) and 3(b). In Figure~5(a), the dotted curve shows the
behavior of the current under the resonance condition (i.e., $\varepsilon=0$). When the energy difference $\varepsilon$ increases, the current
decreases 
[see the dashed and solid curves in Figure~5(a)]. Moreover, within the region $\left\vert eV_{\mathrm{QPC}}\right\vert<\Delta$, the current decreases with increasing $\left\vert eV_{\mathrm{QPC}}\right\vert$ under the off-resonance condition (i.e., $\varepsilon\neq0$). Outside the region, i.e., $\left\vert eV_{\mathrm{QPC}}\right\vert\geq \Delta$, the current varies more nonlinearly with the voltage $%
V_{\mathrm{QPC}}$, as compared with the current under $\varepsilon=0$. Physically, the increase of $\varepsilon$, e.g., from $\varepsilon=0$ to $0.2$ meV, makes the electron tunneling between the two dots more off-resonant, so the current through them decreases\cite{Gurvitz1997PRB}. In addition, the QPC-induced dephasing can yield the broadening of the energy levels at a given nonzero $\varepsilon$, which can further reduce the current\cite{Gurvitz1997PRB}. The effect of dephasing is contrary to the effects of excitation and relaxation which increase the current [see equation~(\ref{current})]. As a result, these two opposite effects compete in the electron-tunneling processes at a given nonzero $\varepsilon$. Within the region $\left\vert eV_{\mathrm{QPC}}\right\vert<\Delta$, the dephasing effect dominates, leading to a decreasing current with the increase of $\left\vert eV_{\mathrm{QPC}}\right\vert$. Outside this region, i.e., $\left\vert eV_{\mathrm{QPC}}\right\vert>\Delta$, both relaxation and excitation processes play an important role, so the current increases with $\left\vert eV_{\mathrm{QPC}}\right\vert$.

The results of the shot noise (i.e., $S/2e$) are shown in Figure~5(b). The shot
noise decreases with the increase of the energy difference $\varepsilon$. At a given nonzero $\varepsilon$, the shot noise decreases with increasing $\left\vert eV_{\mathrm{QPC}}\right\vert$ within the region $\left\vert eV_{\mathrm{QPC}}\right\vert<\Delta$. Outside this region, i.e., $\left\vert eV_{\mathrm{QPC}}\right\vert>\Delta$, the shot noise increases with $\left\vert eV_{\mathrm{QPC}}\right\vert$. These behaviors are similar to those of the current. From both current and shot noise, the Fano factor is straightforwardly obtained, as shown in Figure~5(c). When
$\left\vert eV_{\mathrm{QPC}}\right\vert$ is small, the Fano factor increases with the energy difference $\varepsilon$.
However, for large values of $\left\vert eV_{\mathrm{QPC}}\right\vert$, Fano
factor is suppressed. Moreover, the Fano factor first decreases and then increases with $\left\vert eV_{\mathrm{QPC}}\right\vert$ at a given nonzero $\varepsilon$. These behaviors can be deduced from the current in Figure~5(a) and the shot
noise in Figure~5(b) since $F=S/2eI$. Physically, the off-resonance electron tunneling between two dots is enhanced when increasing $\varepsilon$. This off-resonance makes the two eigenstate channels more asymmetric (i.e., $\Gamma_{R}^{\left( e\right)}\neq \Gamma_{R}^{\left(g \right)}$) around the region $\left\vert eV_{\mathrm{QPC}}\right\vert<\Delta$, and then the Fano factor increases due to dynamical blockaded channels\cite{Belzig2004PRL,Belzig2004EPL,Belzig2005PRB}. For large values of $\left\vert eV_{\mathrm{QPC}}\right\vert$ where backaction becomes strong, 
the decrease of the Fano factor with increasing $\varepsilon$ may be due to the QPC-induced backaction.

The results of the skewness are shown in Figure~5(d). For small values of $\left\vert eV_{\mathrm{QPC}}\right\vert$, the skewness increases with $\varepsilon$. However, the skewness is suppressed when $\left\vert eV_{\mathrm{QPC}}\right\vert$ becomes large. Also, the skewness first decreases and then increases with $\left\vert eV_{\mathrm{QPC}}\right\vert$ at a given nonzero $\varepsilon$. These behaviors are similar to those of the Fano factor, which also indicate that the off-resonance electron tunneling dominates for small values of $\left\vert eV_{\mathrm{QPC}}\right\vert$ and the QPC-induced backaction plays an important role for large values of $\left\vert eV_{\mathrm{QPC}}\right\vert$.

\section*{Discussion}

Note that in the strong Coulomb-blockade regime, we only need to consider the lowest two energy levels of the DQD. In this aspect, it is similar to the single two-level dot in Ref.~45. However, the DQD are different from the single two-level dot in other aspects.
For instance, the DQD provides more controllabilities than a single quantum dot. It can be tuned by not only the two single-dot levels of the DQD but also the interdot tunneling strength via gate voltages. However, in a single quantum dot, only the level spacing can be tuned by the trap potential of the dot. Also, the electron transport through a single quantum dot only involves the tunneling rates $\Gamma_L$ and $\Gamma_R$, while the electron transport through the DQD involves the effective tunneling rates of the two eigenstate channels, i.e., $
\Gamma _{L}^{\left( g\right) }=\alpha ^{2}\Gamma _{L},\Gamma_{L}^{\left(
e\right) }=\beta ^{2}\Gamma _{L},\Gamma _{R}^{\left( g\right) }=\beta
^{2}\Gamma _{R}+\gamma _{\mathrm{ex}}$ and $\Gamma _{R}^{\left( e\right) }=\alpha
^{2}\Gamma _{R}+\gamma _{\mathrm{re}}$ (see Methods), which depend on the energy difference $\varepsilon$ of the two single-dot levels and the tunneling strength
$\Omega$ between these two levels. As seen in Methods, both $\alpha$ and $\beta$ are functions of $\varepsilon$ and $\Omega$. Moreover, in our considered setup, the QPC can induce excitation and relaxation as well as dephasing on the electron that tunnels through the DQD [see the interaction Hamiltonian in the eigenstate basis, i.e., equations~(\ref{eq:interactionH1}), (\ref{eq:interactionH11}) and (\ref{eq:interactionH12}) in Methods]. This is also different from a single quantum dot because the QPC can only induce dephasing on the electron that tunnels through this single dot\cite{Gustavsson2006PRL}.

In summary, we have studied the unavoidable detector-induced backaction on
the counting statistics of a biased DQD. We find that this backaction has
profound effects on the counting statistics, e.g., changing the shot noise
from being sub-Poissonian regime to super-Poissonian, and
changing the skewness from being positive to negative. We also show that
when the energy difference between two single-dot levels of the DQD
increases, both Fano factor and skewness can be either enhanced or suppressed under the detector-induced
backaction. These backaction effects can be experimentally examined by using
the current technologies. Also, our results contribute to possible fine
manipulation of quantum transport processes using the backaction of a charge
detector. 

\section*{Methods}

\noindent \textbf{Quantum dynamics of the DQD.} We derive a master equation
to describe the quantum dynamics of the DQD\cite{You2010PRB}, which is used
to calculate the counting statistics. In the eigenstate basis, the DQD
Hamiltonian can be written as%
\begin{equation}
H_{\mathrm{DQD}}=\frac{\Delta }{2}(|e\rangle \langle e|-|g\rangle \langle
g|),
\end{equation}%
where $\Delta =\sqrt{\varepsilon ^{2}+4\Omega ^{2}}$ is the energy splitting
of the two eigenstates of the DQD given by $|g\rangle =\alpha |1\rangle
-\beta |2\rangle $, and $|e\rangle =\beta |1\rangle +\alpha |2\rangle $, with
$\alpha =\cos ({\theta }/{2})$, $\beta =\sin ({\theta }/{2})$, and $\tan
\theta =2\Omega /\varepsilon $. In the interaction picture with the
unperturbed Hamiltonian $H_{0}\equiv H_{\mathrm{DQD}}+H_{\mathrm{leads}}+H_{%
\mathrm{QPC}}$, the interaction Hamiltonian $H_{\mathrm{I}}\equiv H_{\mathrm{%
T}}+H_{\mathrm{\det }}$ can be written as%
\begin{equation}
H_{\mathrm{\det }}=X\left( t\right) Y\left( t\right) , \label{eq:interactionH1}
\end{equation}%
\begin{eqnarray}
H_{\mathrm{T}}(t)\! &=&\!\sum_{s}\big[c_{ls}^{\dagger }(\alpha
a_{g}e^{-i\Delta t/2}+\beta a_{e}e^{i\Delta t/2})e^{i\omega _{ls}t}+\Upsilon
_{r}^{\dagger }c_{rs}^{\dagger }  \notag \\
&&\times (\alpha a_{e}e^{i\Delta t/2}-\beta a_{g}e^{-i\Delta
t/2t})e^{i\omega _{rs}t}+\mathrm{H.c.}\big],
\end{eqnarray}%
where
\begin{equation}
X\left( t\right) =\sum_{n=1}^{3}U_{n}e^{i\omega _{n}t}, \label{eq:interactionH11}
\end{equation}%
\begin{equation}
Y\left( t\right) =\sum_{kq}V_{kq}^{\dag }\left( t\right) +V_{kq}\left(
t\right) , \label{eq:interactionH12}
\end{equation}%
and $U_{1}=\zeta |e\rangle \langle g|,U_{2}=\zeta |g\rangle \langle
e|,U_{3}=T-\zeta \cos \theta \varrho _{z},$ $\omega _{1}=-\omega _{2}=\Delta
,\omega _{3}=0$, $V_{kq}\left( t\right) =c_{Dq}^{\dag }c_{Sk}e^{-i\left(
\omega _{Sk}-\omega _{Dk}\right) t}.$ $a_{e}$ and $a_{g}$ are annihilation
operators for eigenstates $|e\rangle$ and $|g\rangle$, respectively.

Applying the Born-Markov approximation and tracing over the degrees of
freedom of the QPC, the quantum dynamics of the DQD system in the Schr\"{o}%
dinger picture is governed by the master equation,
\begin{equation}
\dot{\rho}\left( t\right) =-i\left[ H_{\mathrm{DQD}},\rho \left( t\right) %
\right] +\mathcal{L}_{d}\rho \left( t\right) +\mathcal{L}_{T}\rho (t),
\label{eq:masterEquation}
\end{equation}%
with
\begin{eqnarray}
\mathcal{L}_{d}\rho \left( t\right) \!\! &=&\!\!\sum_{i,j=1,i\neq
j}^{3}\left\{ \mathcal{D}\left[ U_{i}\right] \rho \left( t\right) +\mathcal{D%
}\left[ U_{i},U_{j}\right] \rho \left( t\right) \right\} \notag \\
&&\times \lambda \left[ \Theta \left( eV_{\mathrm{QPC}}-\omega _{i}\right)
+\Theta \left( -eV_{\mathrm{QPC}}-\omega _{i}\right) \right] ,
\end{eqnarray}%
\begin{eqnarray}
\mathcal{L}_{T}\rho (t)\!\! &=&\!\!\alpha ^{2}\Gamma _{L}\mathcal{D}%
[a_{g}^{\dagger }]\rho \left( t\right) +\beta ^{2}\Gamma _{L}\mathcal{D}%
[a_{e}^{\dagger }]\rho \left( t\right) \notag \\
&&+\alpha ^{2}\Gamma _{R}\mathcal{D}[a_{e}\Upsilon ^{\dag }]\rho \left(
t\right) +\beta ^{2}\Gamma _{R}\mathcal{D}[a_{g}\Upsilon ^{\dag }]\rho
\left( t\right) \notag \\
&&+\alpha \beta \Gamma _{L}\big\{\left[ a_{e}^{\dagger },\rho \left(
t\right) a_{g}\right] +\left[ a_{g}^{\dagger }\rho \left( t\right) ,a_{e}%
\right] \notag \\
&&+\left[ a_{g}^{\dagger },\rho \left( t\right) a_{e}\right] +\left[
a_{e}^{\dagger }\rho \left( t\right) ,a_{g}\right] \big\} \notag \\
&&-\alpha \beta \Gamma _{R}\big\{\left[ a_{e}\Upsilon ^{\dag },\rho \left(
t\right) a_{g}^{\dagger }\Upsilon \right] +\left[ a_{g}\Upsilon ^{\dag }\rho
\left( t\right) ,a_{e}^{\dagger }\Upsilon \right] \notag \\
&&+\left[ a_{g}^{\dagger }\Upsilon ^{\dag },\rho \left( t\right)
a_{e}^{\dagger }\Upsilon \right] +\left[ a_{e}\Upsilon ^{\dag }\rho \left(
t\right) ,a_{g}^{\dagger }\Upsilon \right] \big\}.
\end{eqnarray}%
Here, $\rho \left( t\right) $ is the reduced density matrix of the DQD
system.
The theta functions $\Theta \left( \pm eV_{\mathrm{QPC}}-\omega _{i}\right)$ appear when tracing over the degrees of freedom of the QPC, i.e., $\int_{-\infty }^{\mu
_{L}}\int_{\mu _{R}}^{+\infty }g_{S}g_{D}d\omega _{k}d\omega
_{q}\int_{0}^{+\infty }d\tau e^{i\omega \tau } e^{-i\left( \omega _{k}-\omega _{q}\right) \tau
}$ $\left\langle c_{Sk}^{\dag }c_{Dq}c_{Dq}^{\dag
}c_{Sk}\right\rangle \!=\! \pi g_{S}g_{D}\Theta \left( eV_{\mathrm{QPC}}-\omega
\right) $, and $\int_{\mu_{L} }^{+\infty }\int_{-\infty}^{\mu _{R}}$ $g_{S}g_{D}d\omega _{k}d\omega_{q}\int_{0}^{+\infty }d\tau e^{i\omega \tau }e^{i\left( \omega _{k}-\omega _{q}\right) \tau}\left\langle c_{Dq}^{\dag}c_{Sk}c_{Sk}^{\dag }c_{Dq}\right\rangle =\pi g_{S}g_{D}\Theta \left( -eV_{\mathrm{QPC}}-\omega
\right) $, with $eV_{\mathrm{QPC}}=\mu_{L}-\mu_{R}$, where $\mu_L$ and $\mu_R$ are the chemical potentials of the source and drain electrodes of the QPC. For $\omega>0$, the QPC-induced excitation occurs when $\left\vert eV_{\mathrm{QPC}}\right\vert>\omega$ [see equation~(\ref{backactionRates1})]. For $\omega<0 $ or $\omega=0$, the QPC-induced excitation or dephasing occurs, respectively [see equations~(\ref{backactionRates2}) and~(\ref{backactionRates3})].
The superoperator $\mathcal{D},$ acting on any single or double operator, is
defined as
\begin{equation}
\mathcal{D}\left[ A\right] \rho \equiv A\rho A^{\dag }-\frac{1}{2}A^{\dag
}A\rho -\frac{1}{2}\rho A^{\dag }A,
\end{equation}%
\begin{equation}
\mathcal{D}\left[ A,B\right] \rho \equiv \frac{1}{2}\big(A\rho B^{\dag
}+B\rho A^{\dag }-B^{\dag }A\rho -\rho A^{\dag }B\big).
\end{equation}%
From equation~(\ref{eq:masterEquation}) and the relations
\begin{eqnarray}
\langle n|\Upsilon _{r}^{\dagger }\rho \Upsilon _{r}|n\rangle &=&\rho
^{(n-1)},~\langle n|\Upsilon _{r}\rho \Upsilon _{r}^{\dagger }|n\rangle
=\rho ^{(n+1)}, \\
\langle n|\Upsilon _{r}^{\dagger }\Upsilon _{r}\rho |n\rangle &=&\rho
^{(n)},~\langle n|\Upsilon _{r}\Upsilon _{r}^{\dagger }\rho |n\rangle =\rho
^{(n)},
\end{eqnarray}%
where $n$ is the number of electrons that have tunneled to the right
electrode of the DQD, we obtain the n-resolved equation of motion for each reduced
density matrix element:
\begin{eqnarray}
\dot{\rho}_{00}^{\left( n\right) } &=&-\Gamma _{L}\rho _{00}^{\left(
n\right) }+\beta ^{2}\Gamma _{R}\rho _{gg}^{\left( n-1\right) }+\alpha
^{2}\Gamma _{R}\rho _{ee}^{\left( n-1\right) }  \notag \\
&&-\alpha \beta \Gamma _{R}(\rho _{eg}^{\left( n-1\right) }+\rho
_{ge}^{\left( n-1\right) }),
\end{eqnarray}%
\begin{eqnarray}
\dot{\rho}_{gg}^{\left( n\right) } &=&\alpha ^{2}\Gamma _{L}\rho
_{00}^{\left( n\right) }-\left( \beta ^{2}\Gamma _{R}+\gamma _{\mathrm{ex}}\right)
\rho _{gg}^{\left( n\right) }+\gamma _{\mathrm{re}}\rho _{ee}^{\left( n\right) }
\notag \\
&&+(\frac{1}{2}\alpha \beta \Gamma _{R}+\eta \gamma _{\mathrm{de}})(\rho
_{eg}^{\left( n\right) }+\rho _{ge}^{\left( n\right) }),
\end{eqnarray}%
\begin{eqnarray}
\dot{\rho}_{ee}^{\left( n\right) } &=&\beta ^{2}\Gamma _{L}\rho
_{00}^{\left( n\right) }+\gamma _{\mathrm{ex}}\rho _{gg}^{\left( n\right) }-(\alpha
^{2}\Gamma _{R}+\gamma _{\mathrm{re}})\rho _{ee}^{\left( n\right) }  \notag \\
&&+(\frac{1}{2}\alpha \beta \Gamma _{R}-\eta \gamma _{\mathrm{de}})(\rho
_{eg}^{\left( n\right) }+\rho _{ge}^{\left( n\right) }),
\end{eqnarray}%
\begin{eqnarray}
\dot{\rho}_{eg}^{\left( n\right) } &=&-i\Delta \rho _{eg}^{\left( n\right)
}+\alpha \beta \Gamma _{L}\rho _{00}^{\left( n\right) }+(\frac{1}{2}\alpha
\beta \Gamma _{R}+\eta \gamma _{\mathrm{ex}})\rho _{gg}^{\left( n\right) }  \notag \\
&&+(\frac{1}{2}\alpha \beta \Gamma _{R}-\eta \gamma _{\mathrm{re}})\rho _{ee}^{\left(
n\right) }-(\frac{1}{2}\Gamma _{R}+2\eta ^{2}\gamma _{\mathrm{de}})\rho _{eg}^{\left(
n\right) }  \notag \\
&&-\frac{1}{2}\left( \gamma _{\mathrm{ex}}+\gamma_{\mathrm{re}}\right) (\rho _{eg}^{\left(
n\right) }-\rho _{ge}^{\left( n\right) }).
\end{eqnarray}%
In particular, we obtain the tunneling rates for two eigenstate channels $%
\Gamma _{L}^{\left( g\right) }=\alpha ^{2}\Gamma _{L},\Gamma_{L}^{\left(
e\right) }=\beta ^{2}\Gamma _{L},\Gamma _{R}^{\left( g\right) }=\beta
^{2}\Gamma _{R}+\gamma _{\mathrm{ex}},$ and $\Gamma _{R}^{\left( e\right) }=\alpha
^{2}\Gamma _{R}+\gamma _{\mathrm{re}}.$ 

\section*{Acknowledgements}

This work is supported by the National Basic Research Program of
China Grant No. 2009CB929302, the National Natural Science
Foundation of China Grant No.
91121015, 
NSF PHY-0925174, and the China Postdoctoral Science Foundation Grant
No. 2012M520146.

\section*{Author Contributions}
Z.Z.L. performed the calculations under the guidance of J.Q.Y., and both C.H.L. and T.Y. also participated in the discussions. All authors contributed to the interpretation of the work and the writing of the manuscript.

%

\clearpage

\noindent
\textbf{Figure~1}~~\textbf{The coupled QDQ-QPC system.} (a) Schematic diagram of a DQD
coupled to two electrodes (S and D) via tunneling barriers. A QPC used for
measuring the DQD electron states yields backaction on the DQD. (b)
Electronic transition between two eigenstates $|g\rangle$ and $|e\rangle$ of
the DQD (with a transition energy $\Delta$) can be induced by the charge
detector QPC. The energy difference $\protect\varepsilon$ between the two
single-dot levels (dashed lines) can be varied by tuning the gate voltages.

\noindent
\textbf{Figure~2}~~\textbf{Current under QPC-induced backaction at the resonance
condition.} (a) Current $I$ versus the QPC bias energy $eV_{\mathrm{QPC}}$
and the tunneling rate $\Gamma_R$. (b) Current $I$ versus the QPC bias
energy $eV_{\mathrm{QPC}}$ for a given tunneling rate $\Gamma _{R}=0.15$
meV. We use typical experimental parameters $\Omega =0.1$ meV and $\Gamma
_{L}=0.05$ meV from Refs.~21 and~22.

\noindent
\textbf{Figure~3}~~\textbf{Fano factor under QPC-induced backaction at the resonance
condition.} (a) Fano factor $F$ versus QPC bias energy $eV_{\mathrm{QPC}}$
and the tunneling rate $\Gamma_R$. (b) Fano factor versus the bias energy $%
eV_{\mathrm{QPC}}$ for a given tunneling rate $\Gamma _{R}=0.15$ meV.
Other parameters are the same as in Figure~2.

\noindent
\textbf{Figure~4}~~\textbf{Skewness under QPC-induced backaction at
the resonance condition.} (a) Skewness $K$ versus QPC bias energy
$eV_{\mathrm{QPC}}$ and the tunneling rate $\Gamma_R$. (b) Skewness versus
the bias energy $eV_{\mathrm{QPC}}$ for a given tunneling rate $\Gamma
_{R}=0.25$ meV. Other parameters are the same as in Figure~2.

\noindent
\textbf{Figure~5}~~\textbf{Current, shot noise, Fano factor and skewness under
QPC-induced backaction at the off-resonance condition.} (a) Current $I$, (b)
shot noise, (c) Fano factor $F$, and (d) skewness $K$ versus
the QPC bias energy $eV_{\mathrm{QPC}}$ for different values of the energy
difference $\protect\varepsilon$ between two single-dot levels of the DQD.
We use a typical experimental parameter $\Gamma_{R}=0.15$ meV. Other
parameters are the same as in Figure~2.

\clearpage
\begin{figure}
\begin{center}
\epsfig{file=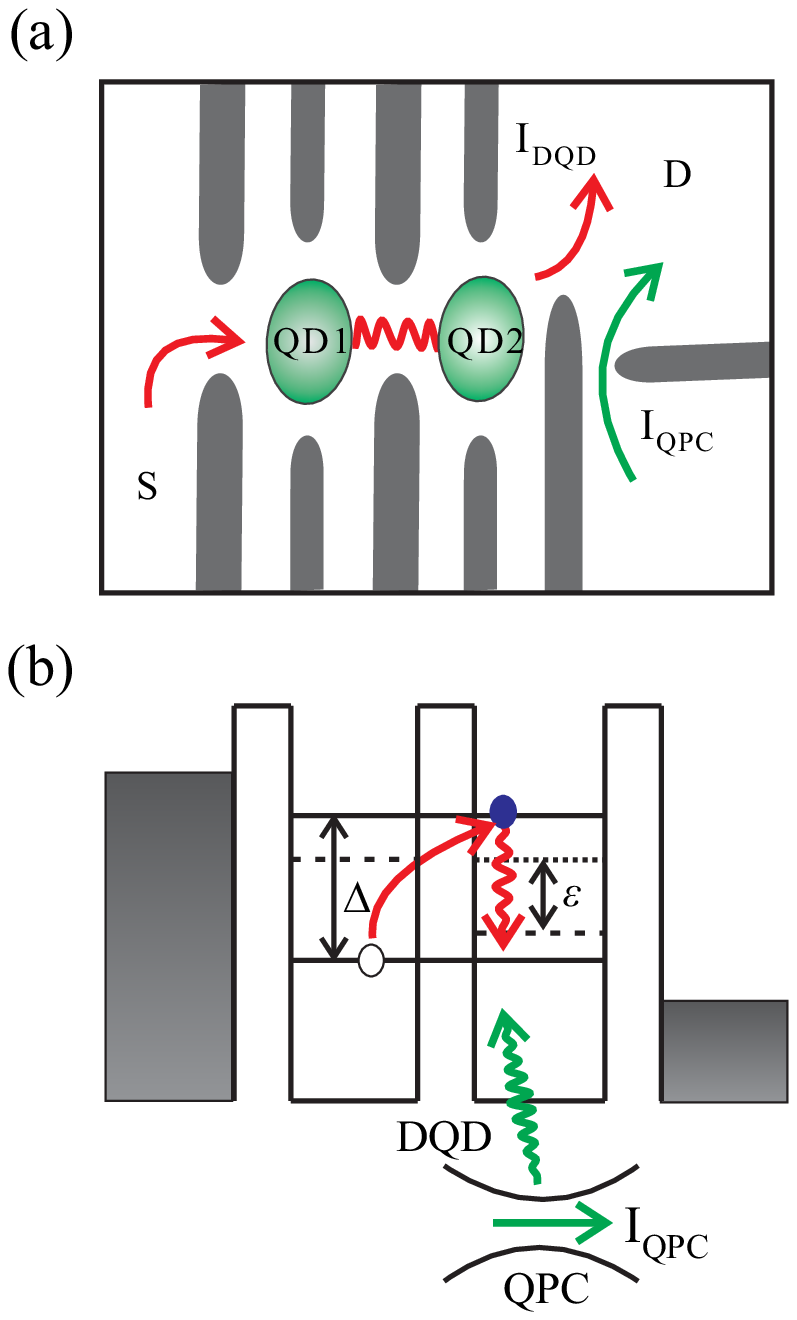,width=0.4\textwidth}
\end{center}
\label{fig:model}
\end{figure}

\clearpage
\begin{figure}
\begin{center}
\epsfig{file=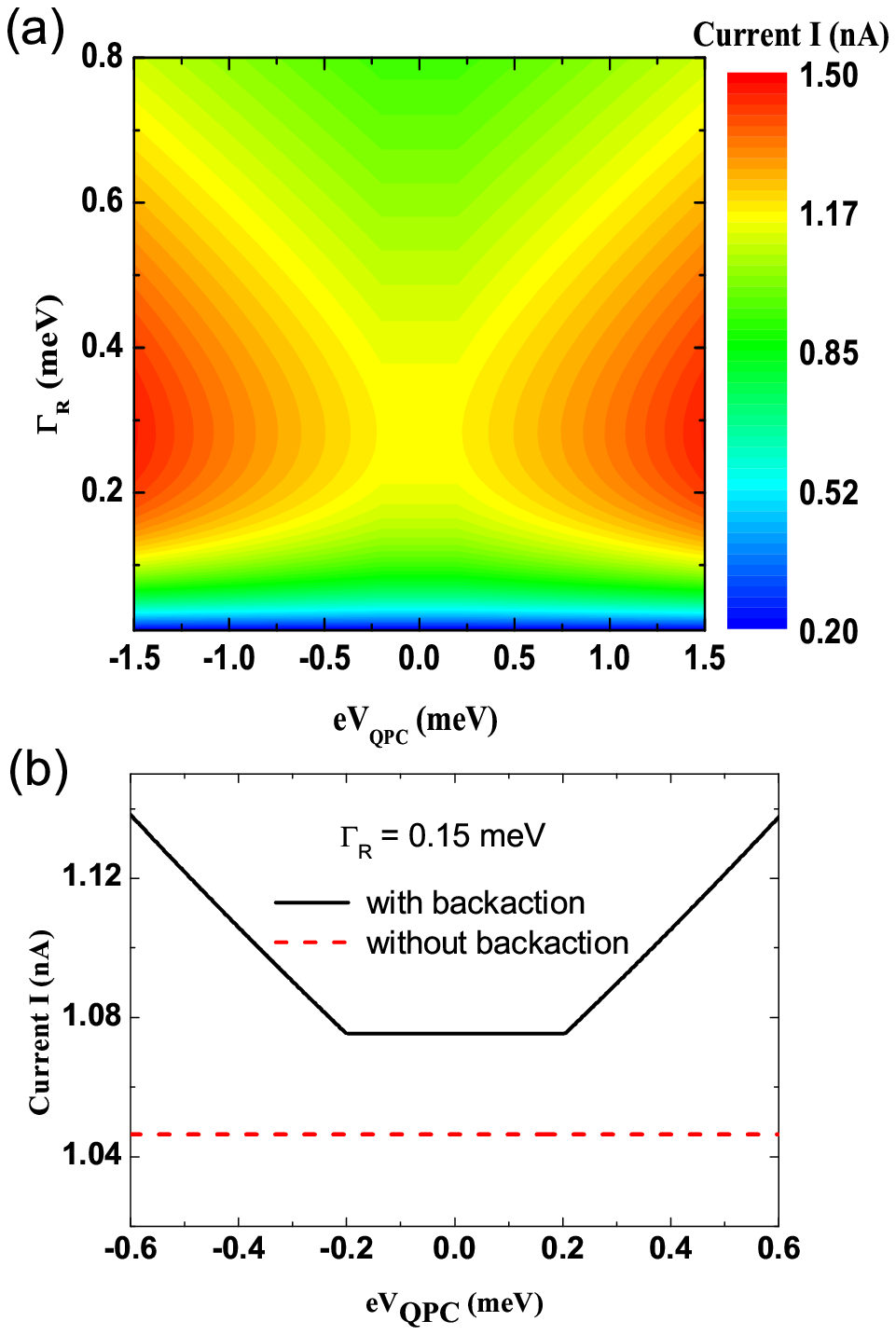,width=0.45%
\textwidth,bbllx=45,bblly=7,bburx=329,bbury=427}
\end{center}
\label{fig_current_backaction}
\end{figure}

\clearpage
\begin{figure}
\begin{center}
\epsfig{file=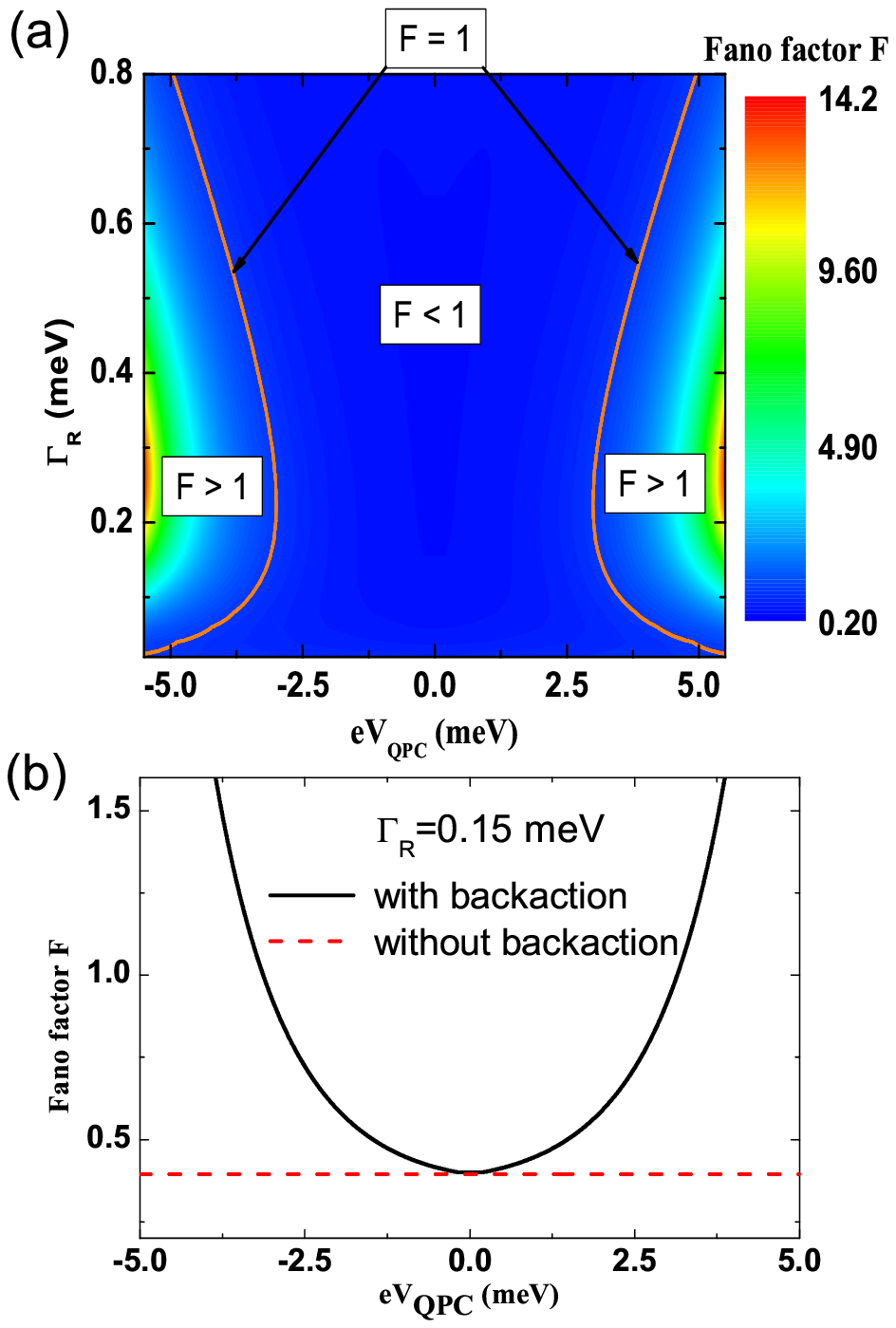,width=0.45%
\textwidth,bbllx=62,bblly=5,bburx=344,bbury=418}
\end{center}
\label{fig_fanofactor_backaction}
\end{figure}

\clearpage
\begin{figure}
\begin{center}
\epsfig{file=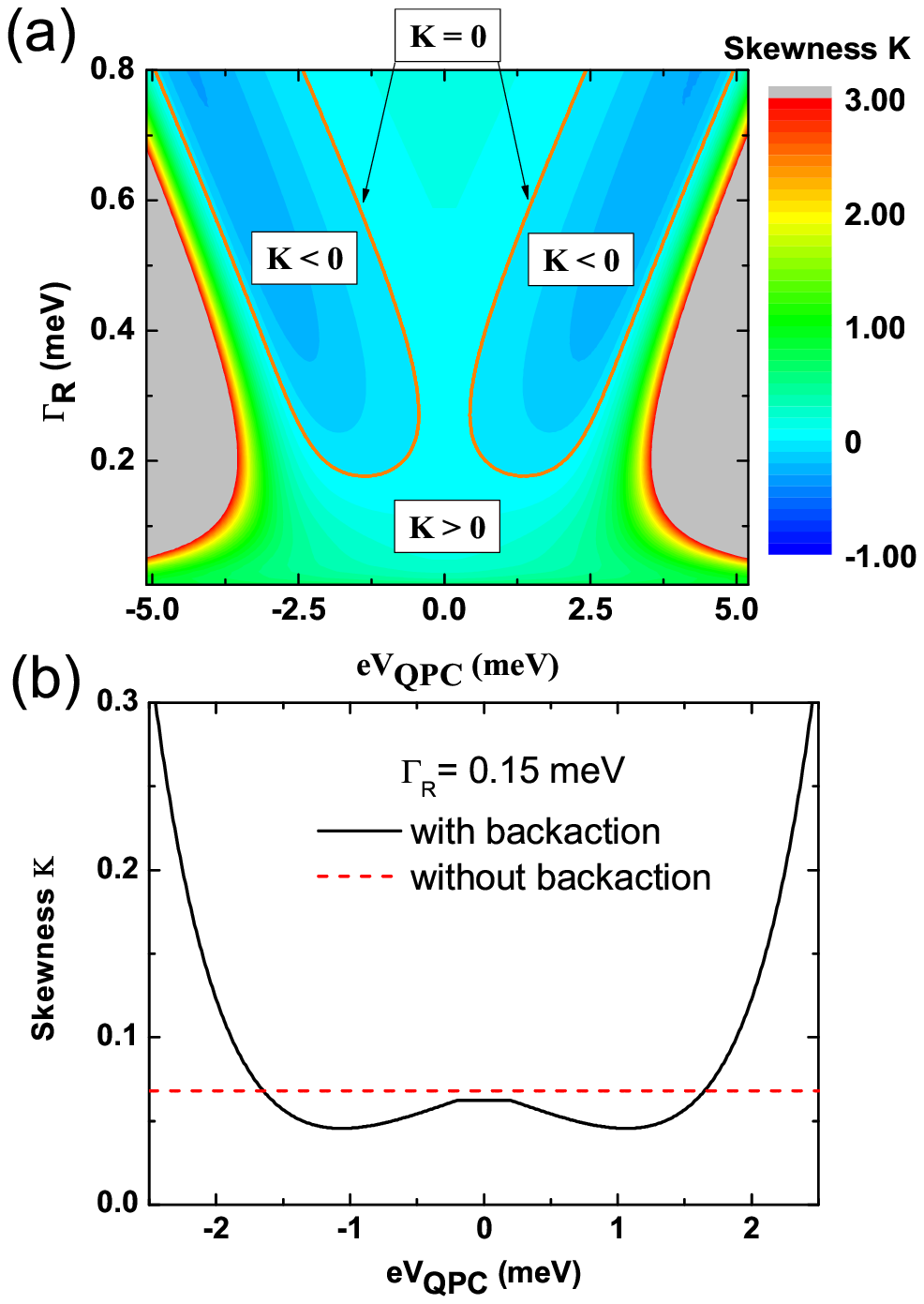,width=0.45%
\textwidth,bbllx=92,bblly=0,bburx=385,bbury=408}
\end{center}
\label{fig_skewness_backaction}
\end{figure}

\clearpage
\begin{figure}
\begin{center}
\epsfig{file=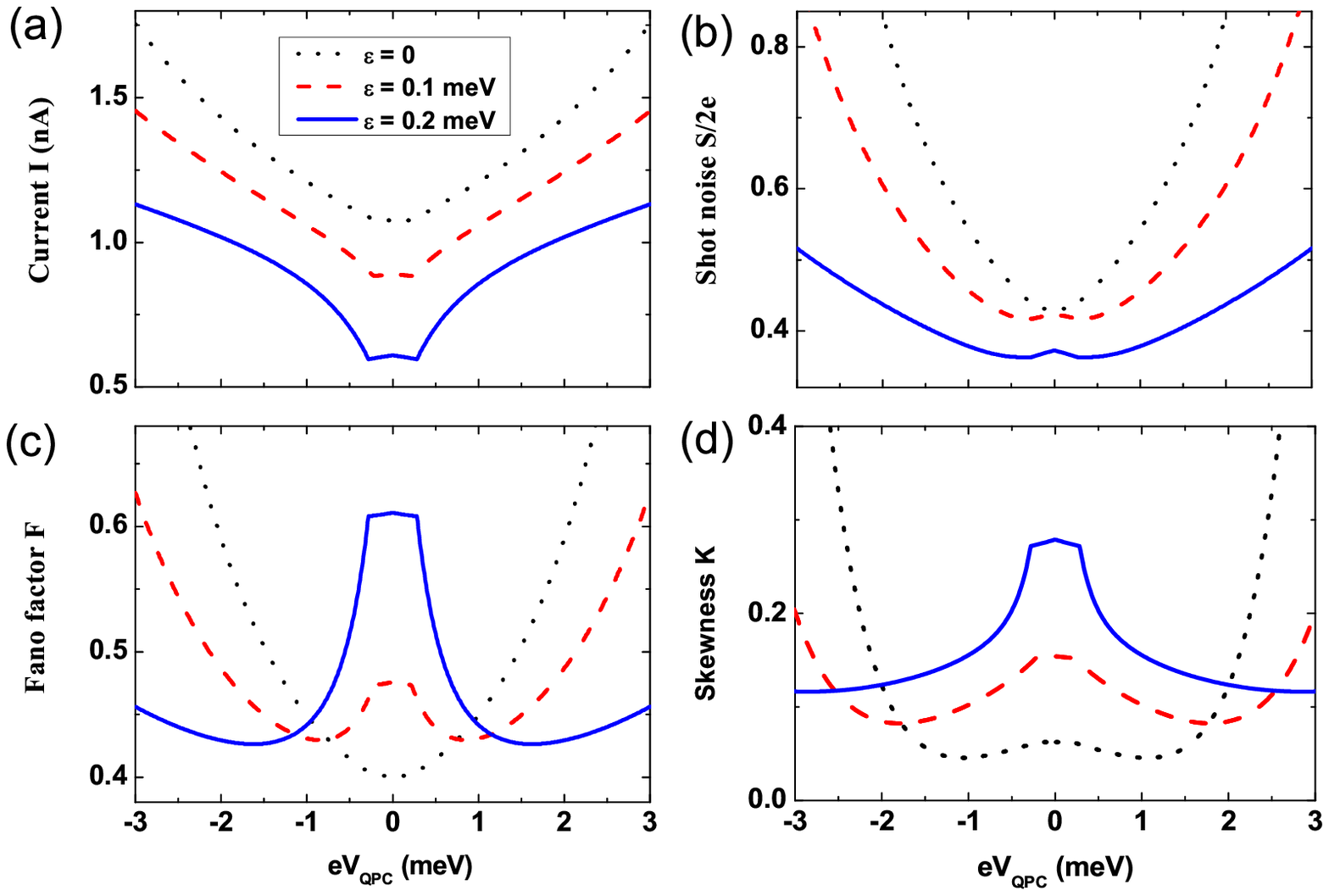,width=0.6%
\textwidth,bbllx=63,bblly=44,bburx=535,bbury=366}
\end{center}
\label{fig_offresonanceCase}
\end{figure}
\clearpage


\clearpage

\begin{thebibliography}{99}
\bibitem{Blanter-Buttiker2000PhysRep} Blanter, Y. \& B\"{u}ttiker, M. Shot noise in mesoscopic conductors. \textit{Phys. Rep.} \textbf{336}, 1-166 (2000).

\bibitem{Nazarov2003} Nazarov, Y. V. \textit{Quantum Noise in Mesoscopic Physics} (Kluwer, Dordrecht, 2003).

\bibitem{Levitov1993JETP} Levitov, L. S. \& Lesovik, G. B. Charge distribution in quantum shot noise. \textit{JETP Lett.} \textbf{\ 58}, 230-235 (1993).

\bibitem{Levitov1996JMP} Levitov, L. S., Lee, H. W. \& Lesovik, G. B. Electron counting statistics and coherent states of electric current. \textit{J. Math. Phys.} \textbf{37}, 4845-4866 (1996).

\bibitem{Rimberg2003Nature} Lu, W., Ji, Z., Pfeiffer, L., West, K. W. \& Rimberg, A. J. Real-time detection of electron tunnelling in a quantum dot. \textit{Nature (London)} \textbf{423}, 422-425 (2003).

\bibitem{Fujisawa2004APL} Fujisawa, T., Hayashi, T., Hirayama, Y., Cheong, H. D. \& Jeong, Y. H. Electron counting of single-electron tunneling current. \textit{Appl. Phys. Lett.} \textbf{84}, 2343-2345 (2004).

\bibitem{BylanderDelsing2005Nature} Bylander, J., Duty, T. \& Delsing, P. Current measurement by real-time counting of single electrons. \textit{Nature (London)} \textbf{434}, 361-364 (2005).


\bibitem{Gustavsson2006PRL} Gustavsson, S. \textit{et al.} Counting statistics of single electron transport in a quantum dot. \textit{Phys. Rev. Lett.} \textbf{96}, 076605 (2006).


\bibitem{Flindt2009PNAS} Flindt, C. \textit{et al.} Universal oscillations in counting statistics. \textit{Proc. Natl. Acad. Sci. USA} \textbf{106}, 10116-10119 (2009).

\bibitem{Gabelli2009PRB} Gabelli, J. \& Reulet, B. Full counting statistics of avalanche transport: An experiment. \textit{Phys. Rev. B} \textbf{80}, 161203(R) (2009).

\bibitem{Fricke2010APL} Fricke, C., Hohls, F., Sethubalasubramanian, N., Fricke, L. \& Haug, R. J. High-order cumulants in the counting statistics of asymmetric quantum dots. \textit{Appl. Phys. Lett.} \textbf{96}, 202103 (2010).

\bibitem{Choi2012APL} Choi, T., Ihn, T., Sch\"{o}n, S. \& Ensslin, K. Counting statistics in an InAs nanowire quantum dot with a vertically coupled charge detector. \textit{Appl. Phys. Lett.} \textbf{100}, 072110 (2012).

\bibitem{Ubbelohde2012} Ubbelohde, N., Fricke, C., Flindt, C., Hohls, F. \& Haug, R. J. Measurement of finite-frequency current statistics in a single electron transistor. \textit{Nature Communications} \textbf{3}, 612 (2012).

\bibitem{Ivanov2012arXiv} Ivanov, D. A. \& Abanov, A. G. Characterizing correlations with full counting statistics: Classical Ising and quantum XY spin chains. \textit{Phys. Rev. E} \textbf{87}, 022114 (2013).

\bibitem{Nazarov2003EPJB} Nazarov, Y. V. \& Kindermann, M. Full counting statistics of a general quantum mechanical variable. \textit{Eur. Phys. J. B} \textbf{35}, 413-420 (2003).

\bibitem{Sukhorukov2007Nphys} Sukhorukov, E. V. \textit{et al.}, Conditional statistics of electron transport in interacting nanoscale conductors. \textit{Nat.\ Phys.} \textbf{3}, 243-247 (2007).

\bibitem{Li-GuopingGuo2012APL} Li, H. O. \textit{et al.} Back-action-induced non-equilibrium effect in electron charge counting statistics. \textit{Appl. Phys. Lett.} \textbf{100}, 092112 (2012).

\bibitem{Wiel2002rmp} van der Wiel, W. G. \textit{et al.} Electron transport through double quantum dots. \textit{Rev. Mod. Phys.} \textbf{75}, 1-22 (2002).

\bibitem{LambertEmaryNori2010prl} Lambert, N., Emary, C., Chen, Y. N. \& Nori, F. Distinguishing quantum and classical transport through nanostructures. \textit{Phys. Rev. Lett.} \textbf{105}, 176801 (2010).

\bibitem{Kiessich2006PRB} Kie{\ss }ich, G., Samuelsson, P., Wacker, A. \& Sch\"{o}ll, E. Counting statistics and decoherence in coupled quantum dots. \textit{Phys. Rev. B} \textbf{73}, 033312 (2006).

\bibitem{Bartholdh2006PRL} Barthold, P., Hohls, F., Maire, N., Pierz, K. \& Haug, R. J. Enhanced shot noise in tunneling through a stack of coupled quantum dots. \textit{Phys. Rev. Lett.} \textbf{96}, 246804 (2006).

\bibitem{Kiessich2007PRL} Kie{\ss }ich, G., Sch\"{o}ll, E., Brandes, T., Hohls, F. \& Haug, R. J. Noise enhancement due to quantum coherence in coupled quantum dots. \textit{Phys. Rev. Lett.} \textbf{99}, 206602 (2007).

\bibitem{Gurvitz1997PRB} Gurvitz, S. A. Measurements with a noninvasive detector and dephasing mechanism. \textit{Phys. Rev. B} \textbf{56}, 15215-15223 (1997).

\bibitem{Korotkov2001PRB} Korotkov, A. N. Selective quantum evolution of a qubit state due to continuous measurement. \textit{\ Phys. Rev. B} \textbf{63}, 115403 (2001).

\bibitem{RuskovKorotkov2003PRB} Ruskov, R. \& Korotkov, A. N. Spectrum of qubit oscillations from generalized Bloch equations. \textit{Phys. Rev. B} \textbf{67}, 075303 (2003).

\bibitem{YoungClerk2010PRL} Young, C. E. \& Clerk, A. A. Inelastic backaction due to quantum point contact charge fluctuations. \textit{Phys. Rev. Lett.} \textbf{104}, 186803 (2010).

\bibitem{You2010PRB} Ouyang, S. H., Lam, C. H. \& You, J. Q. Backaction of a charge detector on a double quantum dot. \textit{Phys. Rev. B} \textbf{81}, 075301 (2010).

\bibitem{Gustavsson2007PRL} Gustavsson, S. \textit{et al.} Frequency-selective single-photon detection using a double quantum dot. \textit{Phys. Rev. Lett.} \textbf{99}, 206804 (2007).

\bibitem{Doiron07} Doiron, C. B., Trauzettel, B. \& Bruder, C. Improved position measurement of nanoelectromechanical systems using cross correlations. \textit{Phys. Rev. B} \textbf{76}, 195312 (2007).

\bibitem{You-Li2012PRB} Li, Z. Z., Ouyang, S. H., Lam, C. H. \& You, J. Q. Probing the quantum behavior of a nanomechanical resonator coupled to a double quantum dot. \textit{Phys. Rev. B} \textbf{85}, 235420 (2012).

\bibitem{Emary2007} Emary, C., Marcos, D., Aguado, R. \& Brandes, T. Frequency-dependent counting statistics in interacting nanoscale conductors. \textit{Phys. Rev. B} \textbf{76}, 161404(R) (2007).

\bibitem{Marcos2010} Marcos, D., Emary, C., Brandes, T. \& Aguado, R. Finite-frequency counting statistics of electron transport: Markovian theory. \textit{New J. Phys.} \textbf{12}, 123009 (2010).

\bibitem{AlbertFlindtButtiker2011prl} Albert, M., Flindt, C. \& B\"{u}ttiker, M. Distributions of waiting times of dynamic single-electron emitters. \textit{Phys. Rev. Lett.} \textbf{107}, 086805 (2011).

\bibitem{Strace-Barrett2004PRL} Stace, T. M. \& Barrett, S. D. Continuous quantum measurement: Inelastic tunneling and lack of current oscillations. \textit{Phys. Rev. Lett.} \textbf{92}, 136802 (2004).

\bibitem{Vandersypen2004APL} Vandersypen, L. M. K. \textit{et al.} Real-time detection of single-electron tunneling using a quantum point contact. \textit{Appl. Phys. Lett.} \textbf{85}, 4394-4396 (2004).

\bibitem{Bagrets-Nazarov2003PRB} Bagrets, D. A. \& Nazarov, Y. V. Full counting statistics of charge transfer in Coulomb blockade systems. \textit{Phys. Rev. B} \textbf{67}, 085316 (2003).

\bibitem{Flindt2005EPL} Flindt, C., Novotn\'{y}, T. \& Jauho, A. P. Full counting statistics of nano-electromechanical systems. \textit{Europhys. Lett.} \textbf{69}, 475-481 (2005).

\bibitem{Flindt2008PRL} Flindt, C., Novotn\'{y}, T., Braggio, A., Sassetti, M. \& Jauho, A. P. Counting statistics of non-Markovian quantum stochastic processes. \textit{Phys. Rev. Lett.} \textbf{100}, 150601 (2008).

\bibitem{Braggio2009JStatMech} Braggio, A., Flindt, C. \& Novotn\'{y}, T. The influence of charge detection on counting statistics. \textit{J. Stat. Mech.} P01048 (2009).

\bibitem{Luo2011JPCM} Luo, J. Y. \textit{et al.} Full counting statistics of level renormalization in electron transport through double quantum dots. \textit{J. Phys.: Condens. Matter} \textbf{23}, 145301 (2011).

\bibitem{Gurvitz1996PRB} Gurvitz, S. A. \& Prager, Y. S. Microscopic derivation of rate equations for quantum transport. \textit{\ Phys. Rev. B} \textbf{53}, 15932-15943 (1996).

\bibitem{Nazarov1996PRB} Nazarov, Y. V. \& Struben, J. J. R. Universal excess noise in resonant tunneling via strongly localized states. \textit{Phys. Rev. B} \textbf{53}, 15466-15468 (1996).

\bibitem{Belzig2004PRL} Cottet, A., Belzig, W. \& Bruder, C. Positive cross correlations in a three-terminal quantum dot with ferromagnetic contacts. \textit{Phys. Rev. Lett.} \textbf{92}, 206801 (2004).

\bibitem{Belzig2004EPL} Cottet, A. \& Belzig, W. Dynamical spin-blockade in a quantum dot with paramagnetic leads. \textit{Europhys. Lett.} \textbf{66}, 405-411 (2004).

\bibitem{Belzig2005PRB} Belzig, W. Full counting statistics of super-Poissonian shot noise in multilevel quantum dots. \textit{Phys. Rev. B} \textbf{71}, 161301 (2005).
%
\end{thebibliography}
\end{document}